\documentclass[12pt,draft,onecolumn,twoside]{IEEEtran}
\usepackage{cite}
\usepackage{amsmath,amssymb,amsfonts,amsthm,mathrsfs}
\usepackage{multirow}
\usepackage{bbm}
\usepackage{algorithmic}
\usepackage{caption}
\usepackage{graphicx}
\usepackage{enumerate}
\usepackage{tikz} 
\usepackage{pgfplots}
\usetikzlibrary{automata, positioning, arrows}
\usepackage{textcomp}
\usepackage{xcolor}
\usepackage{manfnt}
\usepackage{mathtools}
\usepackage{fixmath}
\usepackage{pgf}
\usepackage{xcolor}
\usetikzlibrary{positioning,arrows}
\usepackage{balance} 
\usepackage{csquotes}
\pgfplotsset{compat=1.17}

\def\BibTeX{{\rm B\kern-.05em{\sc i\kern-.025em b}\kern-.08em
    T\kern-.1667em\lower.7ex\hbox{E}\kern-.125emX}}
    
\newcommand{\N}{\mathbb{N}}

\newcommand{\R}{\mathbb{R}}

\newcommand{\C}{\mathbb{C}}

\newcommand{\sA}{\mathcal{A}}

\newcommand{\sE}{\mathcal{E}}

\newcommand{\sM}{\mathcal{M}}
\newcommand{\sP}{\mathcal{P}}

\newcommand{\sL}{\mathcal{L}}
\newcommand{\sN}{\mathcal{N}}

\newcommand{\sX}{\mathcal{X}}

\newcommand{\fT}{\mathfrak{T}}

\theoremstyle{plain}
\newtheorem{thm}{Theorem}
\newtheorem{lem}{Lemma}

\theoremstyle{defn}
\newtheorem{defn}{Definition}

\newtheorem{exmp}{Example}

\theoremstyle{rem}
\newtheorem{rem}{Remark}

\tikzstyle{block} = [draw, rectangle, 
  minimum height=3em, minimum width=4em]

\begin{document}
\title{Characterization of the Arithmetic Complexity of the Secrecy Capacity of Fast-Fading Gaussian Channels}

\author{Holger Boche,~\IEEEmembership{Fellow,~IEEE}, 
    Andrea Grigorescu,~\IEEEmembership{Member,~IEEE}, 
	Rafael F. Schaefer,~\IEEEmembership{Senior Member,~IEEE}, and
	H. Vincent Poor,~\IEEEmembership{Life Fellow,~IEEE} 
	\thanks{H. Boche is with the Chair of Theoretical Information Technology, and the BMBF Research Hub 6G-life, Technical University of Munich, Arcisstr. 21, 80333 M\"unchen, Germany. (E-mail: boche@tum.de). He is also with the Excellence Cluster Cyber Security in the Age of Large-Scale Adversaries (CASA), Ruhr University Bochum.}
    \thanks{A. Grigorescu is with the Chair of Theoretical Information Technology, Technical University of Munich, Arcisstr. 21, 80333 M\"unchen, Germany. (E-mail: andrea.grigorescu@tum.de)}
	\thanks{R. F. Schaefer is with the Chair of Information Theory and Machine Learning, the BMBF Research Hub 6G-life, the Cluster of Excellence ``Centre for Tactile Internet with Human-in-the-Loop (CeTI)'', and the 5G Lab Germany, Technische Universit\"at Dresden, 01062 Dresden, Germany   (E-mail: rafael.schaefer@tu-dresden.de).}
	\thanks{H. Vincent Poor is with the Department of Electrical and Computer Engineering, Princeton University, Princeton, NJ 08544, USA (E-mail: poor@princeton.edu).}         
}
\maketitle

\begin{abstract}
This paper studies the computability of the secrecy capacity of fast-fading wiretap channels from an algorithmic perspective, examining whether it can be computed algorithmically or not. To address this question, the concept of Turing machines is used, which establishes fundamental performance limits of digital computers. It is shown that certain computable continuous fading probability distribution functions yield secrecy capacities that are non-computable numbers. Additionally, we assess the secrecy capacity's classification within the arithmetical hierarchy, revealing the absence of computable achievability and converse bounds.
\end{abstract}
\section{Introduction}

Information-theoretic security was first introduced by Shannon in \cite{shannon1949communication}, where he showed that the one-time pad is information-theoretically secure. In \cite{wyner1975wire}, Wyner characterized the maximal secret communication rate, known as the secrecy capacity, of the wiretap channel. The wiretap channel models a communication scenario involving two legitimate parties, Alice and Bob, and an eavesdropper, Eve. Alice sends messages to Bob through a channel that Bob should decode without error while ensuring the message remains secret from Eve, who intercepts the transmission. The secrecy capacity is the highest communication rate at which Alice and Bob can communicate reliably while preventing Eve from being able to decode the message. Building on this work, Csisz{\'a}r and Körner derived the secrecy capacity for an extended setting, the broadcast channel with confidential messages, in \cite{csiszar1978broadcast}. More recently, \cite{yang2019wiretap} characterized the optimal tradeoff between reliability and secrecy in the finite blocklength regime, providing insights into practical constraints on secure communication. For a comprehensive overview of secure communication models and their extensions, see \cite{liang2009information}.


Ideally, one would develop an algorithm that uses the parameters describing the wiretap channel to compute codes that operate close to the secrecy capacity while ensuring both reliability and secrecy. To achieve this, it is crucial to evaluate how well these codes perform relative to the secrecy capacity, to determine their effectiveness. This evaluation includes developing automated methods for assessing system performance. Therefore, it is essential to establish standards in a machine-readable format and ensure that performance metrics can be computed by digital computers.

 It is often assumed that performance functions, especially those involving entropic quantities such as capacity expressions, are computable. In 1967, methods for constructing both upper and lower bounds for channel reliability functions were introduced in \cite{shannon1967lower}. These techniques were specifically designed to enable the computation of these bounds using digital computers. In 1972, an algorithm to compute the capacity of arbitrary discrete memoryless channels (DMCs) was independently presented in \cite{arimoto1972algorithm} and \cite{blahut1972computation}. In \cite{blahut1972computation}, an analogous algorithm was proposed to compute the rate distortion trade-off of lossy source compression. Typically, channel capacity is expressed through mutual information formulas. It is worth noting that even for the  binary symmetric channel (BSC) with a rational crossover probability, the capacity turns out to be a transcendental number \cite{boche2022algorithmic}. This implies that a precise calculation is not possible, since the computation has to stop after a finite number of steps. Only a suitable approximation of it can be calculated.

The algorithmic computability properties of capacity have been studied for various channels, including finite-state channels (FSCs) \cite{boche2020shannon}, FSCs with feedback \cite{grigorescu2022capacity}, compound channels \cite{boche2020communication}, additive colored Gaussian noise (ACGN) channels \cite{boche2023algorithmic} and fast-fading channels with additive white Gaussian noise (AWGN) \cite{boche2025fast}. For all these channels, it has been shown that their capacities are not computable functions.

While the capacity of the FSC and FSC with feedback is not a computable function, it remains an open question whether their respective capacities are computable as a number. For a channel capacity to be a computable number, it does not need to be a computable function. This means that although a universal algorithm may not exist that takes the channel parameters and a predefined precision as input to compute the capacity to that precision, it is sufficient if, for each channel with computable parameters, the capacity itself is a computable number. In other words, for every channel with computable parameters, there exists an algorithm capable of computing its corresponding capacity. However, for the compound channel, ACGN channel, and fast-fading channel with AWGN, it has been demonstrated that the capacity is not even a computable number. For a clear overview of these results please see Table \ref{table:1}.

\begin{table}
    \begin{center}
    \resizebox{\columnwidth}{!}{%
    \begin{tabular}{||c|c c||} 
        \hline
         Capacity & Computable as a function & Computable as a number\\
        \hline
        Compound Channel & No \cite{boche2020communication} & No \cite{boche2020communication}\\
        ACGN Channel & No \cite{boche2023algorithmic} & No\cite{boche2023algorithmic}\\
        Fast Fading Channel & No\cite{boche2025fast} & No\cite{boche2025fast}\\
        Fast Fading Wiretap Channel & No & No\\
        FSC & No \cite{boche2020shannon} & Open\\
        FSC with Feedback & No \cite{grigorescu2024capacity} & Open\\
        Zero Error for DMC & No \cite{boche2021computability}& Open\\
        \hline
    \end{tabular}
}
    \caption{ Computability of the capacity: current results.}\label{table:1}
\end{center}
\end{table}

In this work, we are interested in algorithmically constructing secrecy codes for the fast-fading Gaussian wiretap channel. Ideally, one could design an algorithm that takes as input the defining parameters of the fast-fading wiretap channel—namely, the fading distribution for the channel with Bob, \( f_b \), and for the channel with Eve, \( f_e \), the corresponding noise powers \( \sigma^2_b \) and \( \sigma^2_e \), the permitted decoding error \( \epsilon \), the power constraint \( P \), and a precision \( M \)—and computes the corresponding code operating at rate \( R \) with \( R \geq C_s(P, f_b, f_e, \sigma^2_b, \sigma^2_e) - \frac{1}{2^M} \), where \( C_s(P, f_b, f_e, \sigma^2_b, \sigma^2_e) \) represents the secrecy capacity of that channel.

This raises the following question:

{\bf Question 1:} \emph{ For a fixed fading channel $(f_b,f_e,\sigma^2_b,\sigma^2_e)$, a fixed error $\epsilon$, and power constraint $P$, is it possible to find an algorithm that gets the precision $M$ as input and computes a code with rate $R$ for the wiretap fast-fading channel, such that $R\geq C(P,f_b,f_e,\sigma^2_b,\sigma^2_e)-\frac{1}{2^M}$ is achieved?}

\begin{figure}
\centering
     \begin{tikzpicture}      [block/.style={draw,minimum width=#1,minimum height=4em},
        block/.default=10em,high/.style={minimum height=3em},auto]

       \node (c) at (-1,0) {$M$};
       \node [block, right=of c] (a) {$\fT_{C_{P,f_b,f_e,\sigma^2_b,\sigma^2_e,\epsilon}}$};
       \node (e) at (7,0) {$(n_M,|\sM_{n_M}|,\epsilon)$};


       \draw[->,draw=black] (c) -- (a);
       \draw[->,draw=black] (a) -- (e);
            
     \end{tikzpicture}
     
     \caption{TM $\fT_{C_{P,f_\alpha,\sigma^2,\epsilon}}$ for the computation of wiretap fast-fading codes operating at $R\geq C(P,f_\alpha,\sigma^2)-\frac{1}{2^M}$.}
	\label{fig:TMCode} 
  \end{figure}
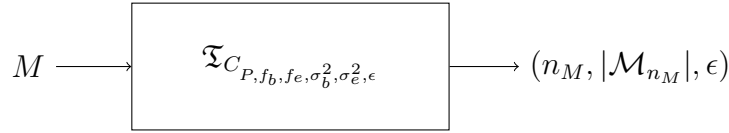

The TM that describes the algorithm of Question 1 is illustrated in Fig. \ref{fig:TMCode}.

The capacity of the ACGN channel has been shown not to be computable as a number, in \cite{boche2023algorithmic}, so an algorithmically capacity-achivable code construction is impossible. However, the arithmetical complexity of the ACGN channel capacity was completly characterized. It has be shown, that the ACGN channel capacity belongs to the set $\Sigma_1$, which consists of all rational numbers for which there exists a computable sequence that converges to the number. This means that for each channel with computable parameters, the capacity is a number for which there exists a monotonically increasing computable sequence $\{R_n\}_{n\in\mathbb{N}}$ of rational numbers that converges to the capacity. In other words, the achievability is computable. Therefore, for each channel, it is possible to find an algorithm that takes an index $n \in \mathbb{N}$ and outputs a rate $R_n$ at which reliable communication is achievable. This prompts the following question:

{\bf Question 2:} \emph{ For a fixed fading channel $(f_b,f_e,\sigma^2_b,\sigma^2_e)$, and power constraint $P$, is it possible to design an algorithm that computes the corresponding sequence of achievable rates $\{R_n\}_{n\in\mathbb{N}}$ that converges to the secrecy capacity?}

On the other hand, for every compound channel with computable parameters, the capacity belongs to the set $\Pi_1$. This implies that for the capacity of any computable compound channel, the converse is computable, i.e., there exists a monotonically decreasing computable sequence $\{U_n\}_{n\in\mathbb{N}}$ of computable numbers that converges to the compound capacity. Consequently, for each computable compound channel, there exists an algorithm that that gets $n\in\N$ as input and provides a rate $U_n$, for which it is impossible to find codes that enable reliable communication at that rate. This reaises the following question:

{\bf Question 3:} \emph{ For a fixed fading channel $(f_b,f_e,\sigma^2_b,\sigma^2_e)$, and power constraint $P$, is it possible to design an algorithm that computes the corresponding sequence of upper bounds $\{U_n\}_{n\in\mathbb{N}}$ on the secrecy capacity, such that the sequence converges to the actual secrecy capacity?}

To study channel capacities algorithmically, we utilize the concept of a Turing machine \cite{turing1938computable,weihrauch2000computable}. This model simulates computation by manipulating symbols on tape based on predefined rules. Turing machines, able to perform any algorithm, are unlimited in computational complexity and error-free operation, setting fundamental limits for current digital computers. They handle all tasks computable on classical machines and match the von Neumann architecture's capabilities without hardware constraints \cite{godel1930vollstandigkeit,godel1934undecidable,kleene1952introduction,minsky1961recursive}. Additionally, we explore non-computability degrees using the arithmetical hierarchy of real numbers \cite{zheng2001arithmetical}.

In this work, we aim to analyze the secrecy capacity of the fast-fading wiretap channel from an algorithmic perspective. Specifically, we investigate whether there exists an algorithm that takes the describing parameters of the fast-fading wiretap channel as input and computes its secrecy capacity to any desired precision—i.e., whether the secrecy capacity is a computable function of the channel parameters. Additionally, we narrow the scope of this question to examine whether the secrecy capacity is, in general, a computable number. Furthermore, we characterize the arithmetical complexity of the secrecy capacity as a number. In particular, we provide a sharp characterization of the degree of non-computability of the secrecy capacity. This characterization has interesting practical implications for computing capacity-achieving code constructions (i.e., achievability bounds) and for the computation of upper bounds (i.e., converse bounds).


\section{Channel Model}\label{sec:chanmod}
We consider the time discrete fast-fading wiretap channel with AWGN. The wiretap channel consists of a sender, Alice, a legitimate receiver, Bob, and an eavesdropper, Eve. The channel input-output relation between Alice, Bob and Eve, at time instance $i\in\{1,\dots,n\}$ is given by
\begin{align*}
	y(i) = \alpha_1(i) x(i) + n_1(i)\\
    z(i) = \alpha_2(i) x(i) + n_2(i)
\end{align*}
where $x(i)$ is the input symbol, and $,y(i),z(i)$ are the output symbols, $\alpha_1(i)$ and $\alpha_2(i)$ are the fading coefficient for the channel from Alice to Bob and Alice to Eve respectively, $n_1(i)\sim \mathcal{N}(0,\sigma^2_1)$ and  $n_2(i)\sim \mathcal{N}(0,\sigma^2_2)$ are AWGN with noise power $\sigma^2_1$ and $\sigma^2_1$ at time instance $i$.

 The fading processes $\{\alpha_1(i)\}$ and $\{\alpha_2(i)\}$ are stationary ergodic processes. Both $\{\alpha_1(i)\}$, $\{\alpha_2(i)\}$ and the additive noise $\{z_i\}$ are i.i.d. across different time instances. The fading coefficients are distributed according to the probability density functions (pdf) $f_{\alpha_1}$ and $f_{\alpha_2}$ respectively, i.e., 
\begin{align*}
    \Pr[|A_1|^2\leq \lambda]&=\int_{-\lambda}^\lambda f_{\alpha_1}(\alpha_1)\,d\alpha_1\\
    \Pr[|A_2|^2\leq \lambda]&=\int_{-\lambda}^\lambda f_{\alpha_2}(\alpha_2)\,d\alpha_2
\end{align*}
where $f_{\alpha_1}$ and $f_{\alpha_2}$ are absolute continuous pdfs. 

 We represent the wiretap fading channel with AWGN by its parameters $(f_{\alpha_1},f_{\alpha_1},\sigma_1^2,\sigma_2^2)$. 

 Let us consider the case where only the receiver has access to the channel state information (CSI). At the same time, the transmitter only knows the statistical properties of the channel but does not have any knowledge about the actual channel realization. The transmission is subject to an average power constraint $P$, i.e., 
\begin{equation*}
    \frac{1}{n}\sum_i^n |x_\ell(i)|^2\leq  P
\end{equation*}
where $\ell$ represents the $\ell$-th message of a codebook.

Next, we introduce the definition of a code and the achievable rate for the fading channel.

\begin{defn}\label{def:code}
    An $(n,|\sM_n|,\epsilon)$\emph{-code} of length $n\in\N$ and size $|\sM_n|$ for the wiretap fading channel consists of an encoder $f_n\colon \{1,\dots,|\sM_n|\}\rightarrow \sP(\sX^n)$ that maps the message $\ell\in\{1,\dots,|\sM_n|\}$ into a codeword $x_\ell^n\in\C^n$, and a decoder $d_n\colon \C^n\rightarrow \{1,\dots,|\sM_n|\}$ that maps the received sequence $y^n\in\C^n$ into a message $\hat{\ell}\in \{1,\dots,|\sM_n|\}$ with an average error probability of 
\begin{equation*}
	\sE=\mathbb{P}\{\hat{\ell}\neq \ell\}.
\end{equation*}
\end{defn}

\begin{defn}\label{def:achievableRate}
    A real number $R$ is called an \emph{achievable secrecy rate} for the wiretap fading channel $(f_{\alpha_1},f_{\alpha_1},\sigma_1^2,\sigma_2^2)$ if for every $\epsilon,\delta>0$ there is an $n_0=n_0(\epsilon,\delta)$ such that for every $n\geq n_0$ there exists an $(n,|\sM_n|,\epsilon)$-code with
    \begin{align*}
        \frac{\log |\sM_n|}{n}&\geq R-\delta \\
        \sE&\leq \epsilon\\
        I(M_n;Z^n)&\leq \epsilon.
    \end{align*}
\end{defn}
This leads to the definition of the capacity for the fast-fading channel.
\begin{defn}
    The supremum of all achievable secrecy rates for the wiretap fast-fading channel $(f_{\alpha_1},f_{\alpha_1},\sigma_1^2,\sigma_2^2)$ subject to an average power constraint $P$ is called \emph{secrecy capacity} and it is denoted by $C_S(P,f_{\alpha_1},f_{\alpha_1},\sigma_1^2,\sigma_2^2) $.
\end{defn}

The capacity formula of the wiretap fast-fading channel with CSI at the receiver only is given in the following Theorem:
\begin{thm}[\cite{lin2016fast}]
   The secrecy capacity of a fast-fading channel $(f_{\alpha_1},f_{\alpha_1},\sigma_1^2,\sigma_2^2)$ subject to an average power constraint $P$ is  

\begin{align}
    C_S(P,f_{\alpha_1},f_{\alpha_1},\sigma_1^2,\sigma_2^2) &= \int_{-\infty}^{+\infty} \log \Big(1+\frac{P|\alpha_1|^2}{\sigma_1^2}\Big)f_{\alpha_1}(\alpha_1)\,d\alpha_1 \nonumber\\
    &-\int_{-\infty}^{+\infty} \log \Big(1+\frac{P|\alpha_2|^2}{\sigma_2^2}\Big)f_{\alpha_2}(\alpha_2)\,d\alpha_2\nonumber
\end{align}
    
     where the optimal input distribution is $\sN(0,P)$.
\end{thm}

\begin{rem}

    The proof of the theorem, as presented in \cite{lin2016fast}, is not constructive, i.e., it does not provide an algorithm for generating secrecy capacity-achieving codes. More specifically, it would be desirable to find an algorithm that takes $(P, f_{\alpha_1}, f_{\alpha_2}, \sigma_1^2, \sigma_2^2)$, a blocklength $n \in \mathbb{N}$, and an error $\epsilon \in (0,1)$ as input parameters and computes the corresponding $(n, |\mathcal{M}_n|, \epsilon)$-code. This code operates at a secrecy rate $R_n = \frac{1}{n} \log |\mathcal{M}_n|$, where $R_n$ is the $n$-th element of the blocklength-dependent  achievable secrecy rate sequence $\{R_n\}_{n \in \mathbb{N}}$. Furthermore, the sequence of secrecy rates should converge asymptotically to $C_S(P, f_{\alpha_1}, f_{\alpha_2}, \sigma_1^2, \sigma_2^2)$, i.e., $\lim_{n \to \infty} R_n = C_S(P, f_{\alpha_1}, f_{\alpha_2}, \sigma_1^2, \sigma_2^2)$.
\end{rem}

\section{Computability Framework}\label{sec:computability_numbers}
This section outlines computability theory, essential for understanding algorithmic limits and capabilities. Turing introduced computability and computable real numbers, describing those precisely determinable by Turing machines \cite{turing1936computable, turing1938computable}. We also examine non-computability using the arithmetical hierarchy of real numbers, which impacts the construction of achievability results and converse bounds \cite{zheng2001arithmetical}.


\subsection{Computable Numbers}
A sequence of rational numbers $\{r_n\}_{n\in\N}$ is called a \emph{computable sequence} if there exist recursive functions $a,b,s:\N\rightarrow\N$ with $b(n)\neq 0$ for all $n\in\N$ and
\begin{equation*}
	r_n= (-1)^{s(n)}\frac{a(n)}{b(n)}, \qquad n\in\N.
\end{equation*}

A real number $x$ is said to be computable if there exists a computable sequence of rational numbers $\{r_n\}_{n\in\N}$, such that
\begin{equation}
	\label{eq:computability_comp2}
	|x-r_n|<2^{-n}
\end{equation}
for all $n\in\N$. This means that the computable real number $x$ is completely characterized by the recursive functions $a,b,s:\N\rightarrow\N$. It has the representation $(a,b,s)$, which we also write as $x\sim (a,b,s)$. It is clear that this representation must not be unique and that there might be other recursive functions $a',b',s':\N\rightarrow\N$ which characterize $x$, i.e., $x\sim (a',b',s')$.

We denote the set of computable real numbers by $\R_c$, which encompasses numbers that can be computed to any desired precision by a Turing machine in a finite number of steps.

\subsection{Computable Sequences}
In this paper, we extensively work with the concepts of computable sequences of computable numbers and effective convergence. These fundamental notions provide a framework for understanding how sequences can be computed and how they converge in a manner that is algorithmically verifiable.

\begin{defn}
	\label{def:computableseq}
	{A sequence of real numbers $\{x_n\}_{n\in\N}$ is \emph{computable} (as a sequence) if there is a computable double sequence of rationals $\{r_{m,n}\}_{m,n\in\N^2}$ such that 
	\begin{equation*}
		|r_{m,n}-x_n|\leq 2^{-m}
	\end{equation*}
	for all $m\in\N$ and $n\in\N$.}
\end{defn}

In what follows, we introduce the concept of Cauchy sequences and their subset, effectively convergent Cauchy sequences.
\begin{defn}
	A sequence $\{x_n\}_{n\in\N}$ is called a \emph{Cauchy sequence} if for every $\epsilon>0$, there is a $n_0\in\N$ such that for every $m,n>n_0$ it holds that
	\begin{equation*}
		|x_n-x_m|<\epsilon.
	\end{equation*}
\end{defn}

\begin{defn}
	A Cauchy sequence $\{x_n\}_{n\in\N}$ is said to converge \emph{effectively} if  there is a recursive function $e\colon\N\times\N\rightarrow\N$ such that for all $n,N\in\N$ it holds that
	\begin{equation*}
	k\geq e(n,N) \quad \text{ implies } |x_k-x_n|\leq 2^{-N}
	\end{equation*}
\end{defn}

\begin{rem}
Let $\{x_n\}_{n\in\N}$ and $\{y_n\}_{n\in\N}$ be computable sequences of real number. Then the following sequence are also computable:
	\begin{equation*}
		x_n\pm y_n,\text{ }x_n y_n,\text{ } x_n/ y_n\text{ } (y_n\neq 0 \text{ for all } n),\text{ } \exp x_n,\text{ } \log x_n \text{ }(x_n>0 \text{ for all } n).
	\end{equation*}
\end{rem}

\subsection{Computable Functions}
Next we introduce two notions of computable functions.
\begin{defn}
	\label{def:borel}
	A function {$f_c:\R_c\rightarrow\R_c$} is called \emph{Borel-Turing computable} if there is an algorithm (or TM) that transforms each given representation $(a,b,s)$ of a computable real number $x$ into a corresponding representation for the computable real number {$f_c(x)$}.
\end{defn}

Next we introduce the notion of \emph{computable continuous functions} \cite[Def.~A]{pour2017computability}. For this, let $\mathbb{I}_c$ denote a computable interval, i.e., $\mathbb{I}_c=[a,b]$ with $a,b\in\R_c$.

\begin{defn}[\cite{pour2017computability}]
	\label{def:compcont}
	Let $\mathbb{I}_c\subset\R_c$ be a computable interval. A function {$f_c:\mathbb{I}_c\rightarrow\R_c$} is called  \emph{computable continuous} if:
	\begin{enumerate}
		\item {$f_c$} is \emph{sequentially computable}, i.e., {$f_c$} maps every computable sequence $\{x_n\}_{n\in\N}$ of real numbers $x_n\in\mathbb{I}_c$ into a computable sequence {$\{f_c(x_n)\}_{n\in\N}$} of real numbers,
		\item {$f_c$} is \emph{effectively uniformly continuous}, i.e., there is a recursive function $d:\N\rightarrow\N$ such that for all $x,y\in\mathbb{I}_c$ and all $N\in\N$ with
		\begin{equation*}
			\|x-y\|\leq\frac{1}{d(N)} \Rightarrow
			{|f_c(x)-f_c(y)|\leq\frac{1}{2^N}.}
		\end{equation*}
	\end{enumerate}
\end{defn}

We continue with the definition of effective convergence.
\begin{defn}
A sequence $\{f_n\}_{k\in\N}$ of computable functions {$f_n:\mathbb{I}_c\rightarrow\R_c$} converges \emph{effectively} to a function $f$, if there exists a recursive function $e\colon\N\rightarrow\N$ such that for all $N\in\N$ it holds that
	\begin{equation*}
		k\geq e(N) \quad \text{implies}\quad |f_n(x)-f(x)|\leq 2^{-N}\quad \text{for all }x\in\mathbb{I}_c.
	\end{equation*}
\end{defn}

In this work, we consider functions in $\sL^1(\R)$, the space of Lebesgue integrable functions, and introduce the notion of computability in $\sL^1(\R)$.

\begin{defn}
    Let $B\in\R_c$ be arbitrary but fixed. A function $f\in\sL^1(\R)$ is called computable in $\sL^1(\R)$ if there exists a computable sequence of computable functions $\{f_n\}_{n\in\N}$ in $\sL^1(\R)$ and a recursive function $e\colon\N\rightarrow\N$ such that for all $M\in\N$ we have
    \begin{equation*}
        |f-f_n|_1\leq \frac{1}{2^M} \quad \text{for all } n\geq e(M).
    \end{equation*}
    
\end{defn}

To describe a non-computable number, we need the notions of recursive enumerable and recursive sets.

\begin{defn}
	\label{def:recursive}
	A set $\sA\subset\N$ is called \emph{recursive} if there exists a computable function $f$ such that $f(x)=1$ if $x\in\sA$ and $f(x)=0$ if $x\notin\sA$. 
\end{defn}

\begin{defn}
	\label{def:recursiveenumerable}
	A set $\sA\subset\N$ is \emph{recursively enumerable} if there exists a recursive function whose domain is exactly $\sA$.
\end{defn}

We have the following properties \cite{soare1978recursively}:
\begin{itemize}
	\item $\sA$ is recursive is equivalent to: $\sA$ is recursively enumerable and $\sA^c$ is recursively enumerable.
	\item There exist recursively enumerable sets $\sA\subset\N$ that are not recursive, i.e., $\sA^c$ is not recursively enumerable. This means there are no computable, i.e., recursive functions $f:\N\rightarrow\sA^c$ with $[f(\N)]=\{m\in\N\colon \exists n \in\N\;\; \text{with}\;\; f(n)=m\}=\sA^c$.
\end{itemize}
Next we introduce an example of a non-computable number based on recursive enumerable non-recursive sets.
\begin{exmp}\label{ex:noncomp}
     Let $\varphi_\sA\colon\N\rightarrow \N$ be a one-to-one recursive function generating a recursively enumerable non-recursive set $\sA$. The number \begin{equation*}
        x_\sA =\sum_{\ell=1}^\infty \frac{1}{2^{\varphi_\sA(l)}} 
    \end{equation*}
    is a non-computable number, since the sequence $\{\sum_{\ell=1}^k \frac{1}{2^{\varphi_\sA(l)}}\}_{k\in\N}$ converges non-effectively to its limit $x_\sA$.
\end{exmp}
%
%

\subsection{Arithmetical Hierarchy}
In this section, we introduce the arithmetical hierarchy of real numbers, as defined by Zheng in \cite{zheng2001arithmetical}. The classification of a real number within this hierarchy is determined by the logical structure used to define it.

The arithmetical hierarchy is closely related to the concept of Turing degrees, which classify the complexity of decision problems for sets of natural numbers (see \cite{post1944recursively,kleene1954upper}). In the context of real numbers, this connection becomes evident when we consider that real numbers can often be represented as the limits of sequences of rational numbers, where each rational is expressed as the quotient of two natural numbers. The complexity of these sequences is linked to the underlying sets of natural numbers that encode them. These sets are frequently interpreted as decision problems, where the task is to determine whether a given number belongs to a specific set. For non-computable sets of natural numbers, their classification in the arithmetical hierarchy reflects their degree of unsolvability, similar to how Turing degrees classify the computational difficulty of problems. 

\begin{defn}[\cite{zheng2001arithmetical}] \label{def:sigma}
		For \(n\in\N_+\), consider \(m^n = (m_1,\ldots,m_n) \in\N^n\). Then, the sets \(\Sigma_n \subsetneq \R\), \(\Pi_n \subsetneq \R\) and
		\(\Delta_n \subsetneq \R\) are defined as follows:
		\begin{itemize}
			\item A number \(x_*\in\R\) satisfies \(x_*\in\Sigma_n\) if there exists an \(n\)-fold computable sequence \((r_{m^n})_{m^n\in\N^n}\)
				of rational numbers such that 
				\begin{align*}
					x_* = \sup_{m_1\in\N}\inf_{m_2\in\N}\sup_{m_3\in\N}\ldots\Theta_{m_n\in\N}~ (r_{m^n})
				\end{align*}
				holds true, where \enquote{\(\Theta\)} is replaced by \enquote{\(\inf\)} if \(n\) is even and by \enquote{\(\sup\)} if \(n\) is odd.
			\item A number \(x_*\in\R\) satisfies \(x_*\in\Pi_n\) if there exists an \(n\)-fold computable sequence \((r_{m^n})_{m^n\in\N^n}\)
				of rational numbers such that 
				\begin{align*}
					x_* = \inf_{m_1\in\N}\sup_{m_2\in\N}\inf_{m_3\in\N}\ldots\Theta_{m_n\in\N}~ (r_{m^n})
				\end{align*}
				holds true, where \enquote{\(\Theta\)} is replaced by \enquote{\(\sup\)} if \(n\) is even and by \enquote{\(\inf\)} if \(n\) is odd.
			\item A number \(x_*\in\R\) satisfies \(x_*\in\Delta_n\) if satisfies both \(x_*\in\Sigma_n\) and \(x_*\in\Pi_n\). Hence, we have
				\(\Delta_n = \Sigma_n \cap \Pi_n\). 
		\end{itemize}
	\end{defn}

 	\begin{rem}
		A real number \(x\) is computable if and only if it satisfies both \(x \in \Pi_1\) and \(x \in \Sigma_1\). Hence, we have \(\R_c = \Delta_1\). 
	\end{rem}

\begin{rem}
    In information theory, if a capacity expression is classified as belonging to $\Sigma_1$, this implies the existence of algorithms that can generate a sequence $\{R_n\}_{n \in \mathbb{N}}$ of computable lower bounds that converge to the capacity. In practical terms, this means that the achievability bounds are computable. Specifically, there exists an algorithm that takes the channel parameters and a natural number $n$ as inputs and outputs the $n$-th element $R_n$ of the achievable rate sequence. Similarly, if a capacity expression belongs to $\Pi_1$, this implies that the converse bounds are computable. In this case, there exists an algorithm that takes the channel parameters and a natural number $n$ as inputs and outputs the $n$-th element of an upper-bound sequence $\{U_n\}_{n\in\N}$ that converges to the capacity.
\end{rem}

\begin{rem}
The set \emph{$\Sigma_1$} is the set of numbers $\underline{x}\in\R$, such that there is a computable sequence of rational numbers $\{\mu_n\}_{n\in\N}$ with $\mu_n\leq\mu_{n+1}$ and $\lim_{n\rightarrow\infty}\mu_n=\underline{x}$. Conversely, The set \emph{$\Pi_1$} is the set of numbers $\bar{x}\in\R$, such that there is a computable sequence of rational numbers $\{\mu_n\}_{n\in\N}$ with $\mu_n\geq\mu_{n+1}$ and $\lim_{n\rightarrow\infty}\mu_n=\bar{x}$.
\end{rem}

\begin{defn}\label{def:delta2}
The set \emph{$\underline{\Delta}_2$} is the set of numbers $x\in\R$, for which there exist $a,b\in\Sigma_1$ such that 
    \begin{equation*}
        x=a-b.
    \end{equation*}
\end{defn}
Note that there are $x\in\underline{\Delta}_2$ that cannot be represented as the limit of a monotonically decreasing computable sequence of reals or as the limit of a monotonically increasing computable sequence of reals, i.e., $x\notin\Sigma_1$ and $x\notin\Pi_1$.

Let $x\in\underline{\Delta}_2$ with $x>0$ be arbitrary, i.e., there are $x_A,x_B\in\Sigma_1$ with $x_A>x_B$ such that 
\begin{equation*}
    x = x_A-x_B
\end{equation*}
We can assume w.l.g. that $x_B>0$, otherwise we could choose a computable number $c$ with $x_B+c>0$ such that 
\begin{align*}
    x&= x_A-c+c-x_B\\
    &=x_A+c -(c+x_B).
\end{align*}
It also holds that $x_A+c\in\Sigma_1$ and $c+x_B\in\Sigma_1$. If we add $c$ to an arbitrary number $x\in\underline{\Delta}_2$ with $x\notin\Sigma_1$ and $x\notin\Pi_1$, then we have that $x+c\in\underline{\Delta}_2$ and $x+c\notin\Sigma_1$ and $x+c\notin\Pi_1$. This means that the logical complexity is determined by the number $x$. Hence, the degree of non-computability or non-approximability of the number $x+c$ is determined by $x$.

\section{Computability of Wiretap Fading Capacity}\label{sec:computabilityWiretap}

In this section, we study the computability of the secrecy capacity of the fast-fading wiretap channel with AWGN. Specifically, we show that the secrecy capacity is not a computable number and, consequently, not a computable function of the channel parameters. Furthermore, we explore its arithmetical complexity and establish that the secrecy capacity falls outside both $\Sigma_1$ and $\Pi_1$ categories, indicating that neither achievability nor converse bounds are computable.

\begin{thm}\label{thm:1}
    Let $x \in \underline{\Delta}_2$ with $x > 0$ and $x \notin \Sigma_1$ as well as $x \notin \Pi_1$ be arbitrary. Let $\sigma_1^2 > 0$, $\sigma_2^2 > 0$, and $\sigma_1^2, \sigma_2^2 \in \R_c$ be arbitrary but fixed. Then there exist continuous $L^1(\R)$-computable pdfs $f_{\alpha_1}$ and $f_{\alpha_2}$ such that for every computable $P > 0$, there is a computable number $u(P)$ satisfying 
    \begin{equation*} 
        C_S(P, f_{\alpha_1}, f_{\alpha_2}, \sigma_1^2, \sigma_2^2) = x + u(P). 
    \end{equation*} 
 Therefore, for all $P \in \R_c$, $P > 0$, we have $C_S(P, f_{\alpha_1}, f_{\alpha_2}, \sigma_1^2, \sigma_2^2) \in \underline{\Delta}_2$, while also satisfying $C_S(P, f{\alpha_1}, f_{\alpha_2}, \sigma_1^2, \sigma_2^2) \notin \Sigma_1$ and $C_S(P, f_{\alpha_1}, f_{\alpha_2}, \sigma_1^2, \sigma_2^2) \notin \Pi_1$.
\end{thm}

\begin{proof}
    Let $x\in\underline{\Delta}_2$ be arbitrary with $x>0$. Let $x_1,x_2\in\Sigma_1$ with $x_1,x_2>0$ and $x=x_1-x_2$ arbitrary.

    Assume we can build two computable continuous pdfs $f_1,f_2$ with $f_1(\alpha_1)=0$ for  $|\alpha_1|\leq 1$, and  $f_2(\alpha_2)=0$ for $|\alpha_2|\leq 1$, both $L^1(\R)$-computable functions, such that it holds that
    \begin{equation*}
        \int_{-\infty}^{\infty} \log (\alpha_1)^2f_1(\alpha_1) \,d\alpha_1=x_1    
    \end{equation*}
    and
    \begin{equation*}
        \int_{-\infty}^{\infty} \log (\alpha_2)^2f_2(\alpha_2) \,d\alpha_2=x_2.    
    \end{equation*}   

    Then we have that for all $P>0$, $P\in\R_c$ and for all $\sigma^2_1,\sigma^2_2>0$ with $\sigma^2_1,\sigma^2_2\in\R_c$ we have that 
    \begin{align*}
        \int_{-\infty}^{\infty} \log (1+\frac{P\alpha_1^2}{\sigma_1^2})^2f_1(\alpha_1) \,d\alpha_1 - \int_{-\infty}^{\infty} \log (\alpha_1)^2f_1(\alpha_1) \,d\alpha_1
        = z_1(P,\sigma^2_1)\in\R_c.
    \end{align*}
    The fact that $z_1(P,\sigma^2_1)\in\R_c$ follows from Lemma \ref{lem:2}.
    The number $z_2(P,\sigma^2_2)$ is defined in the same way as $z_1(P,\sigma^2_1)$, where the defining parameters correspond to $f_2,\sigma^2_2$ and it also holds that $z_2(P,\sigma^2_2)\in\R_c$.

    Furthermore we have that
    \begin{align*}
         \int_{-\infty}^{\infty} &\log (\frac{P\alpha_1^2}{\sigma^2_1})^2f_1(\alpha_1) \,d\alpha_1 =  \int_{-\infty}^{\infty} \log (\alpha_1^2)f_1(\alpha_1) \,d\alpha_1\\
         &\quad+ \log\frac{P}{\sigma^2}\int_{-\infty}^{\infty} f_1(\alpha_1) \,d\alpha_1\\
         &= \int_{-\infty}^{\infty} \log (\alpha_1^2)f_1(\alpha_1) \,d\alpha_1+ \log\frac{P}{\sigma^2}.
    \end{align*}

    It also holds that 
    \begin{align*}
        \int_{-\infty}^{\infty} &\log (1+\frac{P\alpha_1^2}{\sigma_1^2})^2f_1(\alpha_1) \,d\alpha_1 - \int_{-\infty}^{\infty} \log (1+\frac{P\alpha_2^2}{\sigma_2^2})^2f_2(\alpha_2) \,d\alpha_2\\
        &=  \int_{-\infty}^{\infty} \log (\alpha_1^2)f_1(\alpha_1) \,d\alpha_1 -\int_{-\infty}^{\infty} \log (\alpha_2^2)f_2(\alpha_2) \,d\alpha_2\\
         &\quad + z_1(P,\sigma^2_1) - z_2(P,\sigma^2_2)\\
         &= x_1-x_2 + \tilde{z}(P,\sigma^2_1,\sigma^2_2)
    \end{align*}
    where $\tilde{z}(P,\sigma^2_1,\sigma^2_2) = z_1(P,\sigma^2_1) - z_2(P,\sigma^2_2)\in\R_c$ and hence 
    \begin{equation}\label{eq:result}
        \int_{-\infty}^{\infty} \log (1+\frac{P\alpha_1^2}{\sigma_1^2})^2f_1(\alpha_1) \,d\alpha_1 - \int_{-\infty}^{\infty} \log (1+\frac{P\alpha_2^2}{\sigma_2^2})^2f_2(\alpha_2) \,d\alpha_2\in\R_c
    \end{equation}

    As a next step, we construct the pdfs $f_1,f_2$ for which \eqref{eq:result} holds.
    Let $\{a_n\}_{n\in\N}$ be computable sequence of computable numbers that is monotonically increasing and converges to $x_1$.  Then the sequence $\{a+n-\frac{1}{n}\}_{n\in\N}$ is also a strictly monotonically increasing sequence. 
    Let $a_0 = 0$ and $d_n = a_n - a_{n-1}$ for $n\geq 1$. Then we have that $d_n>0$ for $n\geq 1$ and it holds that
    \begin{align*}
        \sum_{n=1}^M d_n = (a_M-a_{M-1}) + (a_{M-1}-a_{M-2}) + \dots +a_1 - a_0= a_M.
    \end{align*}
    It also holds that $\sum_{n=1}^\infty d_n = x_1$.
    
    We consider the following function 
    \begin{equation*}
        g(\alpha)=\begin{cases}
            0 \quad &\alpha\notin [1,4]\\
            \frac{1}{4}(\alpha-1) \quad &\alpha\in(1,2]\\
            \frac{1}{4} \quad &\alpha\in(2,3)\\
            \frac{1}{4}(4-\alpha) \quad &\alpha\in[3,4].
        \end{cases}
    \end{equation*}
    The function is a non-linear computable $L^1(\R)$-function.

    We next define the sequence $\{\sM(n)\}_{n\in\N}$. For every $n\geq 1$, $n\in\N$ we have
    \begin{equation*}
        \sM(n) = \int_{n+1}^{n+4} \log(\alpha_1^2)g(\alpha_1-n)\,d\alpha_1.
    \end{equation*}

    It also holds 
    \begin{equation}\label{eq:1}
        \int_{1}^{4} \log(\alpha_1+n)^2g(\alpha_1)\,d\alpha_1 = \sM(n).
    \end{equation}
    We have that 
    \begin{align*}
        C(n)&\geq \int_2^3 \log (\alpha_1 +n)^2\,d\alpha_1\\
        &= 2\int_2^3\log(\alpha_1 +n)\,d\alpha_1\\
        &2\int_{n+2}^{n+3}\log(\alpha_1)\,d\alpha_1\\
        &=2(n\log n -1)\Big|_{n+2}^{n+3}\\
        &=2[(n+3)\log(n+3)-(n+2)\log(n+2)]\\
        &>2[(n+3)\log(n+2)-(n+2)\log(n+2)]\\
        &= 2\log(n+2).
    \end{align*}

    From \eqref{eq:1} we have that $\{\sM(n)\}_{n\in\N}$ is a computable squence of computable number. 

    We consider now the following computable sequence 
    \begin{equation}\label{eq:2}
        \{\sum_{n=1}^M \frac{d_n}{\sM(n)}\}_{M\in\N}.
    \end{equation}

    Let us take arbitrary $M_1,M_2\in\N$ such that $M_1>M_2$, then we have that
    \begin{align*}
        \sum_{n=1}^{M_2}\frac{d_n}{\sM(n)}-\sum_{n=1}^{M_1}\frac{d_n}{\sM(n)}&=\sum_{n=1}^{M_2}\frac{d_n}{\sM(n)}\\
        &< \frac{1}{M(M_1+1)}\sum_{n=M+1}^{M_2}d_n\\
        &=\frac{1}{M(M_1+1)}(d_{M_2}-d_{M_1+1})\\
        &<\frac{d_{M_2}}{M(M_1+1)}<\frac{K}{M(M_1+1)}
    \end{align*}
    Where $K$ is the smallest natural number such that $K>x_1$ holds.

    This way we have that $\eqref{eq:2}$ is a Cauchy sequence that converges effectively to the number 
    \begin{equation*}
        z_*=\sum_{n=1}^\infty\frac{d_n}{\sM(n)}.
    \end{equation*}
    Hence $z_*$ is a computable number. 

    We consider the following function 
    \begin{equation*}
        \Psi(u)=2\int_1^4 \log(u+\alpha)g(\alpha)\,d\alpha.
    \end{equation*}
    $\Psi$ is a strictly monotonically increasing computable function in $[0,\infty)$. Hence. there exists an  inverse function $\Phi$ of $\Psi$ on $[0,\infty)$ that is also computable and continuous.
    This way, we have that the following 
    \begin{equation*}
        \{\alpha_n^* = \Phi(z_*\sM(n))\}_{n\in\N}
    \end{equation*}
    is a computable sequence of computable numbers.

    Hence, for $\alpha>0$ the following function
    \begin{equation*}
        f_1(\alpha) = \frac{1}{z_*}\sum_{n=1}^\infty\frac{d_n}{\sM(n)}g(\alpha-\alpha_n^*)
    \end{equation*}
    is a computable function. For $\alpha<0$ we set 
    \begin{equation*}
        f_1(\alpha)\coloneq f_1(|\alpha|).
    \end{equation*}

    For $\alpha>0$ we have
    \begin{align*}
        |f_1(\alpha)-\frac{1}{z_*}\sum_{n=1}^M\frac{d_n}{\sM(n)}g(\alpha-\alpha_n^*)|&\leq \frac{1}{z_*}\sum_{n=1}^\infty\frac{d_n}{\sM(n)}\\
        &\leq \frac{1}{z_*}\frac{1}{M(M+1)}\sum_{n=M+1}^\infty d_n\\
        &\leq \frac{K}{z_*M(M+1)}.
    \end{align*}
    Hence, $\{\frac{1}{z_*}\sum_{n=1}^M\frac{d_n}{\sM(n)}g(\alpha-\alpha_n^*)\}_{M\in\N}$ is a computable sequence of computable functions. This way we have that $f_1$ is a computable continuous function on $[0,\infty)$. Following the same calculation steps, and from $\int_{\infty}^{\infty}g(\alpha)\,d\alpha=\frac{1}{2}$ we can show that $f_1$ is a computable $L^1(\R)$-function.

    It holds that $f_1(\alpha_1)=0$ for $|\alpha_1|\leq 1$ and $f_(\alpha_1)\geq 0$ for $\alpha_1\in\R$. It then follows that 
    \begin{align*}
        \int_{-\infty}^{\infty}f_1(\alpha_1)\,d\alpha_1&=2\int_{0}^{\infty}f_1(\alpha_1)\,d\alpha_1\\
        &2\int_{0}^{\infty} \frac{1}{z_*}\sum_{n=1}^\infty\frac{d_n}{\sM(n)}g(\alpha_1-\alpha_n^*)\,d\alpha_1\\
        &2 \frac{1}{z_*}\sum_{n=1}^\infty\frac{d_n}{\sM(n)}\int_{0}^{\infty}g(\alpha_1-\alpha_n^*)\,d\alpha_1\\
        &2 \frac{1}{z_*}\sum_{n=1}^\infty\frac{d_n}{\sM(n)}=1.
    \end{align*}
    Hence, $f_1$ is a pdf for the fast fading channel.

    Furthermore we have that 
    \begin{align}
        \int_{-\infty}^{\infty}\log(\alpha_1)^2f_1(\alpha_1)\,d\alpha_1 &=2\int_{0}^{\infty}\log(\alpha_1)^2f_1(\alpha_1)\,d\alpha_1 \nonumber\\
        &=2\int_{0}^{\infty}\log(\alpha_1)^2\frac{1}{z_*}\sum_{n=1}^\infty\frac{d_n}{\sM(n)}g(\alpha_1-\alpha_n^*)\,d\alpha_1 \nonumber\\
        &=\frac{2}{z_*}\sum_{n=1}^\infty\frac{d_n}{\sM(n)}\int_{0}^{\infty}\log(\alpha_1)^2g(\alpha_1-\alpha_n^*)\,d\alpha_1\label{eq:33} \\
        &=\frac{2}{z_*}\sum_{n=1}^\infty\frac{d_n}{\sM(n)}\Psi(\alpha_n^*) \nonumber\\
        &=\frac{2}{z_*}\sum_{n=1}^\infty\frac{d_n}{\sM(n)}\frac{z_*\sM(n)}{2}\nonumber\\
        &=\sum_{n=1}^\infty d_n=x_1.
    \end{align}
    $f_2$ can be constructed in a similar way.

    The interchange of integration and summation on the right side of \eqref{eq:33} is allowed, because we have $L_1(\R)$-convergence. The  $L_1(\R)$-convergence is shown in the following way: For $M_2>M_1$, $M_1,M_2\in\N$, we have that
    \begin{align*}
        2&\int_1^\infty\log(\alpha_1)^2|\sum_{n=1}^{M_1}\frac{2d_n}{z_*\sM(n)}g(\alpha_1-\alpha_n^*)-\sum_{n=1}^{M_2}\frac{d_n}{z_*\sM(n)}g(\alpha_1-\alpha_n^*)|\,d\alpha_1\\
        &=2\int_1^\infty\log(\alpha_1)^2|\sum_{n=M_2+1}^{M_1}\frac{2d_n}{z_*\sM(n)}g(\alpha_1-\alpha_n^*)|\,d\alpha_1\\
        &=2\int_1^\infty\log(\alpha_1)^2\sum_{n=M_2+1}^{M_1}\frac{2d_n}{z_*\sM(n)}g(\alpha_1-\alpha_n^*)\,d\alpha_1\\
        &=\frac{4}{z_*}\sum_{n=M_2+1}^{M_1}\frac{d_n}{\sM(n)}\int_1^\infty\log(\alpha_1)^2g(\alpha_1-\alpha_n^*)\,d\alpha_1\\
        &=\frac{4}{z_*}\sum_{n=M_2+1}^{M_1}\frac{d_n}{\sM(n)}\frac{z_*\sM(n)}{2}\\
        &=2\sum_{n=M_2+1}^{M_1}d_n= 2(d_{M_2+1}-d_{M_1})\\
        &2d_{M_2+1}.
    \end{align*}
    This way we have that $\{\log(\cdot)^2\frac{1}{z_*}\sum_{n=1}^\infty\frac{d_n}{\sM(n)}g(\cdot-\alpha_n^*)\}_{M\in\N}$ is a Cauchy sequence in $L^1(\R)$. This sequence converges in the $L^1(\R)$ towards the function $\log(\cdot)f_1(\cdot)$. Which proves \eqref{eq:33}.
\end{proof}
\begin{rem}
   Theorem \ref{thm:1} shows that there exists a fast-fading wiretap channel with computable continuous pdfs, computable noise power, and a computable power constraint such that the secrecy capacity is a non-computable number. Consequently, for this particular channel, it is impossible to design algorithms that take a predefined precision as input and compute the secrecy capacity to that precision. This immediately implies that there is no universal algorithm that, given any computable parameter of the fast-fading wiretap channel, can approximate the secrecy capacity to any desired precision.
\end{rem}

\begin{rem}
    In Theorem \ref{thm:1}, a fast-fading wiretap channel characterized by computable continuous functions $\tilde{f}_{\alpha_1}$ and $\tilde{f}_{\alpha_2}$ was found, such that $C_S(P, \tilde{f}_{\alpha_1}, \tilde{f}_{\alpha_2}, \sigma_1^2, \sigma_2^2) \in \underline{\Delta}_2 \setminus (\Sigma_1 \cup \Pi_1)$.  
    In other words, for this fast-fading wiretap channel, the secrecy capacity is a non-computable number for which neither computable upper bounds nor computable lower bounds exist. Consequently, neither achievable bounds nor converse bounds are computable.
\end{rem}



\begin{thm}\label{thm:3}
    Let $f_{\alpha_{1}}, f_{\alpha_{2}}$ be $L^1(\R)$-computable continuous functions. Let $P>0$, $P\in\R_c$, $\sigma^2_1,\sigma^2_2\in \R_c$ with $\sigma^2_1>0$ and $\sigma ^2_2>0$ arbitrary. Then we have that 
    \begin{equation*}
        C_S(P,f_{\alpha_1},f_{\alpha_1},\sigma_1^2,\sigma_2^2)\in\underline{\Delta}_2.
    \end{equation*}
\end{thm}

\begin{proof}
    We consider the following monotonically increasing computable sequences $\{a_n\}_{n\in\N}$ and $\{b_n\}_{n\in\N}$ of computable numbers, where
    \begin{align*}
        a_n&=\int_{-n}^n\log(1+\frac{P\alpha^2_1}{\sigma^2_1})f_1(\alpha_1)\,d\alpha_1\\
        b_n&=\int_{-n}^n\log(1+\frac{P\alpha^2_2}{\sigma^2_2})f_2(\alpha_2)\,d\alpha_2\quad n\in\N.
    \end{align*}
    Here, $f_1$ and $f_2$ are constructed similarly to $f_*$ from Lemma \ref{lem:1}, but with the difference that the arbitrary recursively enumerable, non-recursive set $\sA_1$ used for $f_1$ differs from the set used to construct $f_2$.
    We have that 
    \begin{align*}
        a&=\lim_{n\rightarrow \infty}a_n=\lim_{n\rightarrow \infty}\int_{-n}^n\log(1+\frac{P\alpha^2_1}{\sigma^2_1})f_1(\alpha_1)\,d\alpha_1= \int_{-\infty}^\infty\log(1+\frac{P\alpha^2_1}{\sigma^2_1})f_1(\alpha_1)\,d\alpha_1\\
        b&=\lim_{n\rightarrow \infty}b_n=\lim_{n\rightarrow \infty}\int_{-n}^n\log(1+\frac{P\alpha^2_2}{\sigma^2_2})f_2(\alpha_2)\,d\alpha_2= \int_{-\infty}^\infty\log(1+\frac{P\alpha^2_2}{\sigma^2_2})f_2(\alpha_2)\,d\alpha_2\\
    \end{align*}
    Hence both $a$ and $b$ are both in $\Sigma_1$.
    Since 
    \begin{align*}
        C_S(P,f_{\alpha_1},f_{\alpha_1},\sigma_1^2,\sigma_2^2)&=\int_{-\infty}^\infty\log(1+\frac{P\alpha^2_1}{\sigma^2_1})f_1(\alpha_1)\,d\alpha_1-\int_{-\infty}^\infty\log(1+\frac{P\alpha^2_2}{\sigma^2_2})f_2(\alpha_2)\,d\alpha_2\\
        &=a-b,
    \end{align*}
    from Definition \ref{def:delta2}, we have that $C_S(P,f_{\alpha_1},f_{\alpha_1},\sigma_1^2,\sigma_2^2)\in\underline{\Delta}_2$.
\end{proof}


\begin{rem}
    In Theorem \ref{thm:1}, we show the existence of a fast-fading wiretap channel with computable parameters such that its secrecy capacity belongs to $\underline{\Delta}_2$. Theorem \ref{thm:3} extends this result by showing that for fast-fading wiretap channels with computable, continuous pdfs that are computable in $L^1(\R)$, their secrecy capacity is an element of $\underline{\Delta}_2$. This means that the secrecy capacity can always be represented as the difference between two reals in $\Sigma_1$. However, since the secrecy capacity is expressed as the difference between two $\Sigma_1$ reals, it may not belong to either $\Sigma_1$ or $\Pi_1$, indicating the non-existence of computable upper and lower bounds.
\end{rem}

\begin{rem}
 If the responses to Question~2 were positive, it would then be possible to find an algorithm that generates at least a monotonically increasing sequence $\{R_n\}_{n\in\mathbb{N}}$ of achievable rates converging to the secrecy capacity. However, since the secrecy capacity of some fast-fading wiretap channels is not a computable number, it would be impossible to algorithmically determine how far each $R_n$ is from the secrecy capacity for these channels. This issue arises in the capacity of fast-fading channels \cite{boche2025fast} and ACGN channels \cite{boche2023algorithmic}.
 
 Similarly, if the response to Question 3 were positive, it would be possible to find an algorithm that generates monotonically decreasing sequences $\{U_n\}_{n\in\mathbb{N}}$ of upper bounds converging to the secrecy capacity, i.e., converses, such as for the compound channel \cite{boche2020communication}. However, since the secrecy capacity of some fast-fading wiretap channels is not a computable number, it would be impossible to algorithmically determine how far each $U_n$ is from the secrecy capacity for these channels.
 
 If the responses of both, Question~2 and Question~3, were positive, then we could be able to determine the blocklenght performance, i.e., for some channels it would be possible to derive an algorithm that gets the blocklenght $n \in\N$ as input an computes the respective achievable $R_n$ and converse $U_n$. Howerver, this would not hold for all fast-fading channels with computable parameters, such an algorithm does neither for fast-fading channels, nor ACGN channels nor compound channels, see \ref{table:1}.
 
 If the responses to both Question~2 and Question~3 were positive, we could determine the blocklength performance; that is, for some channels, it would be possible to develop an algorithm that takes the blocklength \(n \in \mathbb{N}\) as input and computes the respective achievable rate \(R_n\) and converse \(U_n\). However, this would not hold for all fast-fading channels with computable parameters. Creating such an algorithm is also impossible for fast-fading channels, ACGN channels, and compound channels, as indicated in Table~\ref{table:1}.
 
Interesting is to see that this is the first result demonstrating that the responses to both Question 2 and Question 3 are negative.
\end{rem}

\section{Conclusions}\label{sec:conclusions}
This paper investigates the algorithmic properties of fast-fading wiretap channels. We show the existence of computable continuous fading pdfs for which the secrecy capacities are non-computable numbers. As a result, it is impossible to design a universal algorithm that takes a fading pdf, noise power, and power constraint as input and computes the secrecy capacity to any desired precision.

Furthermore, we have characterized the degree of non-computability of the secrecy capacity. The secrecy capacity of the fast-fading wiretap channel belongs to \( \underline{\Delta}_2 \), the set of real numbers expressible as the difference between two \( \Sigma_1 \) reals. 

We have also shown the existence of fast-fading wiretap channels with computable continuous pdfs whose secrecy capacity does not belong to \( \Pi_1 \) or \( \Sigma_1 \). For such channels, no computable upper or lower bounds converge to the secrecy capacity, highlighting the increased complexity introduced by secrecy requirements in communication systems. This is the first known result in information theory where the capacity does not belong to either \( \Pi_1 \) or \( \Sigma_1 \). 

By contrast, other capacities shown to be non-computable, such as the compound channel capacity, the ACGN channel capacity, and the zero-error capacity, still belong to the first hierarchy. Specifically, the compound capacity belongs to \( \Pi_1 \), while the ACGN and zero-error capacities belong to \( \Sigma_1 \). For capacities in \( \Sigma_1 \), the achievability bounds are computable, and for those in \( \Pi_1 \), the converse bounds are computable. In the case of the fast-fading wiretap channel, however, neither the achievability bounds nor the converse bounds are computable.

\appendix

\begin{lem}\label{lem:1}
    There is a computable continuous non negative even function $f_*$ that fulfills the following conditions 
    \begin{itemize}
        \item $f_*(\alpha)=0$ for $\alpha\in[-1,1]$
        \item $f_*$ is a computable $L^1$-function and $\int_{-\infty}^\infty f_*(\alpha)\,d\alpha=1$
    \end{itemize}
    such that 
    \begin{equation*}
        \int_{-\infty}^{-1} \log |\alpha|^2f_*(\alpha)\,d\alpha + \int_{1}^\infty \log |\alpha|^2f_*(\alpha)\,d\alpha \notin \R_c.
    \end{equation*}
\end{lem}
\begin{proof}
    For $M\in\N$ and $M\geq 4$ we consider the following function
    \begin{equation*}
	g_M(\alpha)=\begin{cases}
	   0 &\text{for } 0\leq\alpha\leq 2\\
	   \frac{\alpha - 2}{3(\log 3)^2} &\text{for } 2<\alpha\leq 3\\
      \frac{1}{\alpha(\log \alpha)^2} &\text{for } 3<\alpha\leq M\\
      \frac{M+1-\alpha}{M(\log M)^2} &\text{for } M<\alpha\leq M+1\\
      0 &\text{for } \alpha > M+1.\\
	\end{cases}
\end{equation*}
 The function is a computable continuous function and it a computable $L^1-$function, It holds that
 \begin{align*}
     \int_3^M &\frac{\log \alpha}{\alpha(\log \alpha)^2} \,d\alpha = \int_3^M \frac{1}{\alpha(\log \alpha)} \,d\alpha \\ 
     &= \int_{\ln 3}^{\ln M} \frac{1}{\beta} \,d\beta=\frac{1}{\ln 2} \log \frac{\ln M}{\ln 3}
 \end{align*}
Hence
\begin{small}
     \begin{align*}
     \int_2^{M+1} &g_M(\alpha)\log \alpha\,d\alpha = \int_0^\infty  g_M(\alpha)\log \alpha\,d\alpha\\
     &= \frac{1}{2}\frac{1}{3 (\log 3)^2}+\frac{1}{\ln 2} \log \frac{\ln M}{\ln 3}+ \frac{1}{2}\frac{1}{M^2(\log M)^2}:=K_M
 \end{align*}
\end{small}

This way we have that $\{K_M\}_{\substack{M\in\N\\ M\geq 4}}$ is a computable sequence of computable numbers.

Let $\sA\subset \N$ be an arbitrary recursive enumerable non-recursive set and $\varphi_\sA$ the generative function of the set $\sA$.

Consider the function $g_*$ given by
\begin{equation*}
    g_*(\alpha)=\sum_{M=4}^\infty \frac{1}{2^{\varphi_
    \sA (M)}}\frac{1}{K_M}g_M(\alpha).
\end{equation*}
For an arbitrary $\hat{M}>4$, we have
\begin{footnotesize}
    \begin{align}
    \Big|&g_*(\alpha)-\sum_{M=4}^{\hat{M}}\frac{1}{2^{\varphi_\sA (M)}}\frac{1}{K_M}g_M(\alpha)\Big|= \Big|\sum_{M=\hat{M}+1}^{\infty}\frac{1}{2^{\varphi_\sA (M)}}\frac{1}{K_M}g_M(\alpha)\Big|\nonumber\\
    &\leq c_1\sum_{M=\hat{M}+1}^{\infty} \frac{1}{2^{\varphi_\sA (M)}}\frac{1}{K_M}< c_1\sum_{M=\hat{M}+1}^{\infty} \frac{1}{2^{\varphi_\sA (M)}}\frac{1}{\frac{1}{\ln 2} \log \frac{\ln M}{\ln 3}}\label{eq:1}\\
    &< c_1\frac{1}{\ln \frac{\ln \hat{M}+1}{\ln 3}}\sum_{M=\hat{M}+1}^{\infty} \frac{1}{2^{\varphi_\sA (M)}}\label{eq:2}\\
    &< c_1\frac{1}{\frac{1}{\ln 2} \ln \frac{\ln \hat{M}+1}{\log 3}}\label{eq:3},
\end{align}
\end{footnotesize}
where \eqref{eq:1} follows from the relation $K_M>\ln \frac{\ln M+1}{\ln 3}$ for every $M\in\N$ and $M\geq 4$ and $c_1 = \frac{1}{3(\log 3)^2}\geq |g_M(\alpha)|$ for $\alpha\geq 0$ and $M\geq 4$.

\eqref{eq:2} holds since the sequence $\{\frac{1}{\ln 2} \ln \frac{\ln M}{\ln 3}\}_{\substack{M\in\N\\ M\geq 4}}$ is strictly monotonically increasing.

\eqref{eq:3} follows from the following relation 
\begin{align*}
    \sum_{M=\hat{M}+1}^{\infty} \frac{1}{2^{\varphi_\sA (M)}}&< \sum_{M=4}^{\infty} \frac{1}{2^{\varphi_\sA (M)}} < \sum_{\ell=1}^{\infty} \frac{1}{2^{\ell}}\\
    &= \sum_{\ell=0}^{\infty} \frac{1}{2^{\ell}}-1= \frac{1}{1-\frac{1}{2}}-1=1.
\end{align*}
This way we have that the computable sequence $\{g_M(\alpha)\}_{\substack{M\in\N\\ M\geq 4}}$ of computable continuous functions $g_M$ converges effectively to $g_*$, where $g_*$ is a computable continuous function. 

Furthermore, we have that 
\begin{footnotesize}
    \begin{align*}
        \int_2^{M+1}g_M(t)\,dt &= \frac{1}{2}\frac{1}{3(\log 3)^2}+\frac{1}{2}\frac{1}{M(\log M)^2}+\int_3^M \frac{1}{\alpha(\log\alpha)^2}\,d\alpha\\
        &=\frac{1}{2}\frac{1}{3(\log 3)^2}+\frac{1}{2}\frac{1}{M(\log M)^2}+\frac{1}{(\ln 2)^2}\int_{\ln 3}^{\ln M} \frac{1}{\beta^2}\,d\beta\\
        &= \frac{1}{2}\frac{1}{3(\log 3)^2}+\frac{1}{2}\frac{1}{M(\log M)^2}+\frac{1}{(\ln 2)^2}(\frac{1}{\ln M}-\frac{1}{\ln 3})\\
        &< \frac{1}{3(\log 3)^2}+ \frac{1}{\log M}
\end{align*}
\end{footnotesize}

This way we have a similar calculation as for the maximization of the norm for $\hat{M}>4$
\begin{align*}
    \int_0^\infty &\Big|g_*(\alpha)-\sum_{M=4}^{\hat{M}}\frac{1}{2^{\varphi_\sA (M)}}\frac{1}{K_M}g_M(\alpha)\Big| \,d\alpha\\
    & \leq \sum_{M=\hat{M}+1}^{\infty}\frac{1}{2^{\varphi_\sA (M)}}\frac{1}{K_M}(\frac{1}{3(\log 3)^2}+ \frac{1}{\log M})\\
    &\leq \frac{1}{\frac{1}{\ln 2} \ln \frac{\ln \hat{M}+1}{\ln 3}}\frac{2}{3(\log 3)^2}
\end{align*}
with $\log \hat{M}>3(\log 3)^2$, i.e., $\hat{M}>3^3(\log 3)$.

Therefore $g_*$ is a computable $L^1$-function with $g_*\geq 0$ for $\alpha\geq 0$. It also holds that 
\begin{equation*}
    \int_0^\infty g_*(\alpha)\,d\alpha = c_2>0
\end{equation*}
and $c_2\in\R_c$. This way we have that
\begin{equation}
    f_*(\alpha)=\begin{cases}
         \frac{g(\alpha)}{2 c_2} \text{ for } \alpha>0\\
        \frac{g(-\alpha)}{2 c_2} \text{ for } \alpha<0
    \end{cases}
\end{equation}
  is a non-negative computable continuous $L^1$-function with
    \begin{equation*}
        \int_{-\infty}^{\infty} f_*(\alpha)\,d\alpha=1.
    \end{equation*}
    The following now holds:
    \begin{align}
        \int_{-\infty}^{-1} &\log|\alpha|^2f_*(\alpha)\,d\alpha + \int_{1}^{\infty} \log|\alpha|^2f_*(\alpha)\,d\alpha\nonumber \\
        & = 2\int_{1}^{\infty} \log|\alpha|^2f_*(\alpha)\,d\alpha=4\int_{1}^{\infty} \log|\alpha| f_*(\alpha)\,d\alpha\nonumber\\
        &= 4 \sum_{M=4}^\infty \frac{1}{2^{\varphi_\sA (M)}}\frac{1}{K_M}\int_{1}^{\infty}\frac{1}{2c_2}g_M(\alpha)\,d\alpha\label{eq:4}\\
         &= \frac{2}{c_2} \sum_{M=4}^\infty \frac{1}{2^{\varphi_\sA (M)}} = \frac{2}{c_2} x_{\sA}\nonumber
    \end{align}
    with $x_{\sA}= \sum_{M=4}^\infty \frac{1}{2^{\varphi_\sA (M)}}$. However $x_{\sA}\notin \R_c$.  In \eqref{eq:4},  we can change the order if the integration and the limit of the series, since the computable sequence $\{g_M\}_{\substack{M\in\N\\ M\geq 4}}$ of computable $L^1$-functions converges effectively to $g_*$. It also holds that $f_*=\frac{1}{2c_2}g_*$. 
\end{proof}

\begin{lem}\label{lem:2}
   There is a computable continuous non-negative even function $f_*$ that fulfills the following conditions 
    \begin{itemize}
        \item $f_*(\alpha)=0$ for $\alpha\in[-1,1]$
        \item $f_*$ is a computable $L^1$-function and $\int_{-\infty}^\infty f_*(\alpha)\,d\alpha=1$
    \end{itemize}
    such that for all $P>0$, $P\in\R_c$, $\sigma^2\in\R_c$ the following holds
    \begin{footnotesize}
        \begin{equation*}
            x_* =  \int_{1}^{\infty} \log\Big(1+\alpha^2\frac{P}{\sigma^2}\Big)f_*(\alpha)\,d\alpha - \int_{1}^{\infty} \log\Big(\alpha^2\frac{P}{\sigma^2}\Big)f_*(\alpha)\,d\alpha
        \end{equation*}      
    \end{footnotesize}
    with $x^*\in\R_c$.
\end{lem}
\begin{proof}
    Let $m>1$, $m\in\N$ be arbitrary. Let
    \begin{footnotesize}
    \begin{equation*}
        x_m = \int_{1}^{m} \log\Big(1+\alpha^2\frac{P}{\sigma^2}\Big)f_*(\alpha)\,d\alpha - \int_{1}^{m} \log\Big(\alpha^2\frac{P}{\sigma^2}\Big)f_*(\alpha)\,d\alpha,
    \end{equation*}
    \end{footnotesize}
    where $\{x_m\}_{m\in\N}$ is a computable sequence of computable numbers.
    Let 
        \begin{align*}
        x_*-x_m &= \int_{m}^{\infty} \log\Big(1+\frac{\sigma^2}{\alpha P}\Big)f_*(\alpha)\,d\alpha\ \\
        &< \log\Big(1+\frac{\sigma^2}{m P}\Big)\int_{m}^{\infty}f_*(\alpha)\,d\alpha\\
        & < \frac{\sigma^2}{m P} < \frac{1}{m} c_3
    \end{align*}

    where $c_3>\frac{\sigma^2}{P}$. This way the sequence $\{x_m\}_{m\in\N}$ converges effectively to $x_*$, i.e., $x_*\in\R_c$.
\end{proof}

\balance
\bibliographystyle{IEEEtran}
\bibliography{IEEEabrv,confs-jrnls,references_coloregaussian}

\begin{thebibliography}{10}
\providecommand{\url}[1]{#1}
\csname url@samestyle\endcsname
\providecommand{\newblock}{\relax}
\providecommand{\bibinfo}[2]{#2}
\providecommand{\BIBentrySTDinterwordspacing}{\spaceskip=0pt\relax}
\providecommand{\BIBentryALTinterwordstretchfactor}{4}
\providecommand{\BIBentryALTinterwordspacing}{\spaceskip=\fontdimen2\font plus
\BIBentryALTinterwordstretchfactor\fontdimen3\font minus
  \fontdimen4\font\relax}
\providecommand{\BIBforeignlanguage}[2]{{%
\expandafter\ifx\csname l@#1\endcsname\relax
\typeout{** WARNING: IEEEtran.bst: No hyphenation pattern has been}%
\typeout{** loaded for the language `#1'. Using the pattern for}%
\typeout{** the default language instead.}%
\else
\language=\csname l@#1\endcsname
\fi
#2}}
\providecommand{\BIBdecl}{\relax}
\BIBdecl

\bibitem{shannon1949communication}
C.~E. Shannon, ``Communication theory of secrecy systems,'' \emph{Bell Syst.
  Tech.~J.}, vol.~28, no.~4, pp. 656--715, Oct. 1949.

\bibitem{wyner1975wire}
A.~D. Wyner, ``The wire-tap channel,'' \emph{Bell Syst. Tech.~J.}, vol.~54,
  no.~8, pp. 1355--1387, Oct. 1975.

\bibitem{csiszar1978broadcast}
I.~Csisz{\'a}r and J.~Korner, ``Broadcast channels with confidential
  messages,'' \emph{{IEEE} Trans. Inf. Theory}, vol.~24, no.~3, pp. 339--348,
  May 1978.

\bibitem{yang2019wiretap}
W.~Yang, R.~F. Schaefer, and H.~V. Poor, ``Wiretap channels: Nonasymptotic
  fundamental limits,'' \emph{{IEEE} Trans. Inf. Theory}, vol.~65, no.~7, pp.
  4069--4093, Jul. 2019.

\bibitem{liang2009information}
Y.~Liang, H.~V. Poor, and S.~Shamai, ``Information theoretic security,''
  \emph{Found. Trends. Commun. Inf. Theory}, vol.~5, no. 4--5, pp. 355--580,
  2009.

\bibitem{shannon1967lower}
C.~E. Shannon, R.~G. Gallager, and E.~R. Berlekamp, ``Lower bounds to error
  probability for coding on discrete memoryless channels. {I},'' \emph{Inf.
  Contr.}, vol.~10, no.~1, pp. 65--103, Jan. 1967.

\bibitem{arimoto1972algorithm}
S.~Arimoto, ``An algorithm for computing the capacity of arbitrary discrete
  memoryless channels,'' \emph{{IEEE} Trans. Inf. Theory}, vol.~18, no.~1, pp.
  14--20, Jan. 1972.

\bibitem{blahut1972computation}
R.~Blahut, ``Computation of channel capacity and rate-distortion functions,''
  \emph{{IEEE} Trans. Inf. Theory}, vol.~18, no.~4, pp. 460--473, Jul. 1972.

\bibitem{boche2022algorithmic}
H.~Boche, R.~F. Schaefer, and H.~V. Poor, ``Algorithmic computability and
  approximability of capacity-achieving input distributions,'' \emph{{IEEE}
  Trans. Inf. Theory}, vol.~69, no.~9, pp. 5449--5462, Sep. 2023.

\bibitem{boche2020shannon}
------, ``Shannon meets {T}uring: Non-computability and non-approximability of
  the finite state channel capacity,'' \emph{Commun. Inf. Syst.}, vol.~20,
  no.~2, pp. 81--116, Nov. 2020.

\bibitem{grigorescu2022capacity}
A.~Grigorescu, H.~Boche, R.~F. Schaefer, and H.~V. Poor, ``Capacity of finite
  state channels with feedback: Algorithmic and optimization theoretic
  properties,'' in \emph{Proc. IEEE Int. Symp. Inf. Theory}, Espoo, Finland,
  Jul. 2022, pp. 498--503.

\bibitem{boche2020communication}
H.~Boche, R.~F. Schaefer, and H.~V. Poor, ``Communication under channel
  uncertainty: An algorithmic perspective and effective construction,''
  \emph{{IEEE} Trans. Signal Process.}, vol.~68, pp. 6224--6239, Oct. 2020.

\bibitem{boche2023algorithmic}
H.~Boche, A.~Grigorescu, R.~F. Schaefer, and H.~V. Poor, ``Algorithmic
  computability of the capacity of {G}aussian channels with colored noise,'' in
  \emph{Proc. IEEE Global Telecommun. Conf.}, Kuala Lumpur, Malaysia, Dec.
  2023, pp. 4375--4380.

\bibitem{boche2025fast}
H.~Boche, A.~Grigorescu, R.~Schaefer, and H.~V. Poor, ``Algorithmic
  characterization of the outage capacity of {G}aussian fading channels,'' in
  \emph{IEEE Int. Conf. Commun.}, in press.

\bibitem{grigorescu2024capacity}
A.~Grigorescu, H.~Boche, R.~F. Schaefer, and H.~V. Poor, ``Capacity of finite
  state channels with feedback: Algorithmic and optimization theoretic
  properties,'' \emph{{IEEE} Trans. Inf. Theory}, vol.~70, no.~8, pp.
  5413--5426, Aug. 2024.

\bibitem{boche2021computability}
H.~Boche and C.~Deppe, ``Computability of the zero-error capacity of noisy
  channels,'' in \emph{Proc. IEEE Inf. Theory Workshop}.\hskip 1em plus 0.5em
  minus 0.4em\relax Kanazawa, Japan: IEEE, 2021, pp. 1--6.

\bibitem{turing1938computable}
A.~M. Turing, ``On computable numbers, with an application to the
  {E}ntscheidungsproblem. {A} correction,'' \emph{Proc. London Math. Soc.},
  vol.~2, no.~43, pp. 544--546, Jan. 1938.

\bibitem{weihrauch2000computable}
K.~Weihrauch, \emph{Computable {A}nalysis: {A}n {I}ntroduction}.\hskip 1em plus
  0.5em minus 0.4em\relax Springer Science \& Business Media, 2000.

\bibitem{godel1930vollstandigkeit}
K.~G{\"o}del, ``Die {V}ollst{\"a}ndigkeit der {A}xiome des logischen
  {F}unktionenkalk{\"u}ls,'' \emph{Monatshefte f{\"u}r Mathematik und Physik},
  vol.~37, no.~1, pp. 349--360, Dec. 1930.

\bibitem{godel1934undecidable}
\BIBentryALTinterwordspacing
------, ``On undecidable propositions of formal mathematical systems,'' May
  1934, {L}ectures at the Institute for Advanced Study, Princeton, NJ.
  [Online]. Available:
  \url{https://albert.ias.edu/entities/publication/9587ae8c-4cc2-424f-ac21-8681a9c19d8a/details}
\BIBentrySTDinterwordspacing

\bibitem{kleene1952introduction}
S.~C. Kleene, \emph{Introduction to {M}etamathematics}.\hskip 1em plus 0.5em
  minus 0.4em\relax Amsterdam : North-Holland Publishing ; Groningen : P.
  Noordhoff N.V., 1952.

\bibitem{minsky1961recursive}
M.~L. Minsky, ``Recursive unsolvability of {P}ost's problem of "tag" and other
  topics in theory of {T}uring machines,'' \emph{Annals Math.}, pp. 437--455,
  Nov. 1961.

\bibitem{zheng2001arithmetical}
X.~Zheng and K.~Weihrauch, ``The arithmetical hierarchy of real numbers,''
  \emph{Math. Log. Quarterly}, vol.~47, no.~1, pp. 51--65, 2001.

\bibitem{lin2016fast}
P.-H. Lin and E.~Jorswieck, ``On the fast fading gaussian wiretap channel with
  statistical channel state information at the transmitter,'' \emph{{IEEE}
  Trans. Inf. Forensics Security}, vol.~11, no.~1, pp. 46--58, Jan. 2016.

\bibitem{turing1936computable}
A.~M. Turing, ``On computable numbers, with an application to the
  {E}ntscheidungsproblem,'' \emph{Proc. London Math. Soc.}, vol.~2, no.~42, pp.
  230--265, Nov. 1936.

\bibitem{pour2017computability}
M.~B. Pour-El and J.~I. Richards, \emph{Computability in {A}nalysis and
  {P}hysics}.\hskip 1em plus 0.5em minus 0.4em\relax Cambridge University
  Press, 2017.

\bibitem{soare1978recursively}
R.~I. Soare, ``Recursively enumerable sets and degrees,'' \emph{Bull. Am. Math.
  Soc.}, vol.~84, no.~6, pp. 1149--1181, 1978.

\bibitem{post1944recursively}
E.~L. Post, ``Recursively enumerable sets of positive integers and their
  decision problems,'' \emph{Bull. Am. Math. Soc.}, vol.~50, no.~5, pp.
  283--316, 1944.

\bibitem{kleene1954upper}
S.~C. Kleene and E.~L. Post, ``The upper semi-lattice of degrees of recursive
  unsolvability,'' \emph{Annals of mathematics}, vol.~59, no.~3, pp. 379--407,
  1954.

\end{thebibliography}
\end{document}